\documentclass[12pt]{article}

\usepackage{amssymb,amsmath,amsthm} 
\usepackage[margin=0.80in]{geometry}
\usepackage{setspace}
\usepackage{graphicx,ctable,booktabs}
\usepackage{amssymb,amsmath,amsthm,amstext,amscd}
\usepackage{hyperref}
\usepackage{amsmath}
\usepackage{algorithm}
\usepackage[noend]{algpseudocode}
\usepackage{mathtools}

\newcommand{\R}{\mathbb{R}}

\usepackage{epstopdf}
\AppendGraphicsExtensions{.tif}
\usepackage{fancyhdr}
\usepackage{color}
\usepackage{graphicx}

\makeatletter
\def\BState{\State\hskip-\ALG@thistlm}
\makeatother
\usepackage{epstopdf}
\usepackage{sectsty}

\sectionfont{\fontsize{12}{0}\selectfont}

\subsectionfont{\fontsize{12}{0}\selectfont}

\title{Analysis and Design of a Passive Switched-Capacitor Matrix Multiplier for Approximate Computing}
\author{Edward Lee \qquad S. Simon Wong \\  Contact: edhlee@stanford.edu \\ Stanford University, Electrical Engineering
Stanford, CA, USA}
\date{\begin{flushleft}
\normalsize{
A switched-capacitor matrix multiplier is presented for approximate computing and machine learning applications. The multiply-and-accumulate operations perform discrete-time charge-domain signal processing using passive switches and $300$aF unit capacitors. The computation is digitized with a $6$b asynchronous SAR. The analyses of incomplete charge accumulation and thermal noise are discussed. The design was fabricated in 40nm CMOS, and experimental measurements of multiplication are illustrated using matched filtering and image convolutions to analyze noise and offset. Two applications are highlighted: 1) energy-efficient feature extraction layer performing both compression and classification in a neural network for an analog front-end and 2) analog acceleration for solving optimization problems that are traditionally performed in the digital domain. The chip obtains measured efficiencies of $8.7$TOPS/W at $1$GHz for the first application and $7.7$TOPS/W at $2.5$GHz for the second application.  
}
\end{flushleft}
}
\usepackage{textcomp}
\begin{document}
\maketitle

\thispagestyle{empty}

\subsection{Keywords}
1) analog computing, 2) approximate computing, 3) neural networks, 4) matched filtering, 5) matrix factorization, 6) switched-capacitor circuits
\section{Introduction}
Matrix multiplication is the fundamental operation $y=Ax$ where $x \in \R^n$ maps to output $y \in \R^m$ by a linear system $A$. It is ubiquitously used in scientific computing, computer graphics, machine learning, real-time signal processing, and optimization. Matrix multiplication in hardware is traditionally realized by multiply-and-accumulate (MAC) units commonly used in general purpose graphics processing units, field programmable gate arrays, and application-specific integrated circuits.  Three important parameters in matrix multiplication are computation speed (e.g. throughput), energy efficiency, and resolution. For example, while high computation speed is of utmost importance for scientific computing and graphics, energy efficiency plays a more significant role for embedded systems. On the other hand, high resolution is used to obtain high accuracies in computational simulations \cite{blackhole}. 

There have been recent works in reduced-precision multiplication for statistical inference systems optimized for energy-efficient operation. These applications operate on inherently noisy data and performs tasks such as classification and recognition that are resilient to low signal-to-noise (SNR). These fundamental ideas are the motivating forces for reduced-precision or approximate computing. Such systems include classification systems for images and audio and supervised training in machine learning \cite{han2013approximate,St.Amant,Vincent,binaryNet,logNet,bishop}. For example, the work of \cite{Vincent} shows that the performance of inference for neural networks is robust at 8b fixed-point. Inference in the context of image recognition entails the prediction result of one image using  programmable weights (e.g. elements in the matrix $A$) that were trained offline. The works of \cite{binaryNet,logNet} show that resolutions for state-of-the-art networks \cite{VGG16} for the ImageNet Challenge \cite{ILSVRC15} can go down to  less than 4b. The ability for these systems to operate at these ultra-low precisions opens up the possibility of scalable CMOS analog signal processing to work in synergy with digital  systems for higher energy efficiency. 

Analog-domain multiply and accumulate operations can also operate on raw analog data obtained from the sensor before digitization. This can alleviate analog-to-digital (A/D) requirements. Traditionally, conventional systems use analog-to-digital matrix multiplication (AD-MM) \cite{lee2015factorization}, which is a common task in modern sensing and communication systems. AD-MM digitizes an analog signal and multiplies the resulting data by a matrix. For example, AD-MM is used in cameras to compress digital data using transform coding and quantization. However, many analog signals are known to have a sparse representation in some basis, which presents an opportunity to reduce the A/D data rate in an analog-to-digital system. For example, \cite{Cite1} designed an analog DCT in an image sensor in order to compress data before digitization. Many recent papers \cite{Cite2,Cite3,Cite4,Cite5,Cite6,Cite7,Cite8,Cite9,AIC1,AIC2,CirculantMatrix} have explored the use of analog MACs to alleviate A/D requirements for AD-MM. 

Analog MAC designs come in many options that are mainly influenced by energy, speed, resolution requirements. Translinear \cite{lu20151}, current-mode \cite{skrzyniarz201624}, and time-based approaches \cite{miyashita2014ldpc, nahlus2014energy} allow for analog computation to meet large dynamic ranges under low supply voltages. However, these approaches are susceptible to variations in process, voltage, and temperature (PVT) for small unit current sources and unit delays. Voltage domain approaches use switches and capacitors to perform the MAC operation. Capacitor sizes dictate the multiplication elements in the matrix $A$ and charge redistribution (either active or passive) performs the accumulation. Switches and capacitors are highly amenable to nanometer CMOS process. For example, switched-capacitor circuits in filters and successive approximation register (SAR) ADCs are designed with signal transfer functions that depend only on ratios of two capacitors and not their absolute capacitance values. Because of this, the size of capacitors and switches can be aggressively scaled to decrease dynamic $CV^2$ energy. Recent experiments on mismatch of metal fringe capacitors \cite{Vaibhav} show that sub-femto fringe capacitors can be realized with around 1\% mismatch (1 std. deviation) in $32$nm. Our implementation uses 300aF capacitors for multiplication. This translates to 1\% gain error (1 std. deviation) for the LSB multiplier, which is well within our 3b multiplication specification. 


Both active and passive switched-capacitor MACs have been implemented in the past for information processing. Active approaches are commonly used when more than 8-9 bits are required \cite{Cite1}. However, since recent machine learning applications can operate with fewer bits, there has been a growing push towards passive switched-capacitor implementations. For example, the work of \cite{zhang201518} embeds charge multiplication in a SAR ADC using switches, capacitors, and digital barrel-shifting in 130nm. Parallel switched-capacitor MACs for neural network applications were proposed in \cite{tsividis1987switched} and \cite{bankman2015passive}. 

Our Switched-Capacitor Matrix Multiplier (SCMM) draws inspiration from these previous works to enable sequential MACs using only passive switches and capacitors to target applications that are resilient to reduced-precision. The system platform for analog matrix multiplication and its various applications are illustrated in Figure 1. MAC operations take elements in matrix $A$, multiply them with elements in vector $x$ in the analog domain, and digitizes the result to produce $y$. The elements of $A$ are stored in memory and are realized using a digital-to-analog converter (DAC) that multiplies $x$. One complete inner product of the $j$-th row of $A$ with the input for example is $y_j=\Sigma_{i=1}^{n} A[j,i]x[i]$, where $n$ is the number of elements in $x$ and also the number of MAC clock cycles. This operation is performed $m$ times for $m$ rows in the matrix $A$.

  Because input data can be either analog or digital, we explore two options for $x$: 1) $x$ is inherently a discrete-time analog voltage $x[i] = V_{\text{in}}[i]$ where $i$ represents its time index and 2) $x$ is a digital vector that then generates the analog voltage $x[i] = V_{\text{in}}[i]$. The first option is aimed at AD-MM for sensing interfaces and communication, where it is possible to not only perform the matrix operation while the data is still analog but also to alleviate A/D requirements after matrix computation. The second option is aimed to accelerate matrix evaluations in machine learning applications where our design interfaces to a digital environment. The final SCMM's energy efficiency of 8.8 (1GHz) and 7.7 TOPS/W (2.5GHz) is computed using the measured total power of the SCMM (input DAC, MAC, SAR, memory, clock, and self-test).

\section{Design of SCMM}

\subsection{Active versus passive MACs} \label{activeVpassive}
There are many conceptual switched-capacitor methods to perform the MAC operation $y_j=\Sigma_{i=1}^{n} A[j,i]x[i]$. Figure 2 highlights two main approaches: (a) active and (b) passive. For both approaches, the input is a voltage $x[i] = V_{\text{in}}[i]$ that is multiplied by $C_1[i]$ during $\phi_1$, where $C_1[i]$ is a capacitive-DAC (CDAC) controlled by the $A[j,i]$ value in memory. During $\phi_2$, the multiplied charge $ V_{\text{in}}[i] C_1[i]$ is redistributed onto the capacitor $C_2$ either by (a) active or (b) passive means. The $\phi_1$ and $\phi_2$ operations are performed $n$ times for each element in the inner-product before quantizing the voltage on $C_2$. However, due to finite gain for active and incomplete charge transfer for passive, the actual inner-product is not ideally $V_{C_2}=\frac{1}{C_2}\Sigma_{i=1}^{n}  V_{\text{in}}[i] C_1[i]$. 

Instead, due to finite amplifier gain $A_0$ in the active approach, the voltage of $C_2$ at the $i$-th cycle is $V_{C_2}[i] = k[i] V_{C_2}[i-1] + \mu[i]k[i] V_{\text{in}}[i]$, where $k[i]= \frac{C_2 (A_{0}+1)}{C_2 (A_{0}+1)+|C_{1}[i]|} $ and $\mu[i]=\frac{C_{1}[i]}{C_2} \frac{A_{0}}{A_{0}+1}$. $k[i]$ is a droop term that represents the fraction of charge on $C_2$ that is leaked away from cycle to cycle (ideally $k[i]=1$), and $\mu[i] k[i]$ contains the voltage gain error at every sample $i$. Since the matrix calculation only cares about the last cycle $i=n$, we can write the output voltage at time $n$ as
\begin{equation}
\label{eq2}
\begin{aligned} 
V_{C_2}[n] & = \sum_{i=1} ^{n} \mu[i] V_{\mathrm{in}}[i]  \prod_{j=i}^{n} k[j]   \\
 & = \begin{bmatrix}
 \mu[1] \prod_{i=1}^{n} k[i] & \mu[2] \prod_{i=2}^{n} k[i]  & \ldots & \mu[n-1] \prod_{i=n-1}^{n} k[i]  & \mu[n]k[n]         \\
     \end{bmatrix} x.
\end{aligned}
\end{equation}

This result can be extended to the passive case in Figure 2(b) where $k[i]= \frac{C_2}{C_2 +|C_{1}[i]|} $ and $\mu[i]=\frac{C_{1}[i]}{C_2}$. Note that in the passive modality, these gain and droop terms only depend on the ratios of capacitors and not the amplifier gain, which is sensitive to nonlinearities. The gain components $\mu[i] k[i]$, which are $\frac{C_1[i]A_{0}}{C_2(A_{0}+1)+|C_1[i]|}$ for active and $\frac{C_1[i]}{C_2+|C_1[i]|}$ for passive, indicate that in order to achieve a more ideal gain of $\frac{C_1[i]}{C_2}$, one has to increase either the DC gain for active or increase the size of $C_2$ for passive. Equating the active and passive gains indicates that in order to achieve the same level of gain error, the active's DC gain must be $\frac{C_2}{|C_1|}+1 \gg 1$. This undesirable attribute implies that the amplifier is unnecessary for low-precisions. At high resolutions on the other hand, the active approach is more compelling since the choice of DC gain offers one extra degree of freedom whereas the passive approach is limited to the ratios of two capacitors. This conclusion also extends to the droop terms $k[i]$. Therefore, this comparison indicates that at least for low-resolution gains, the passive approach can yield the same levels of MAC performance (neglecting noise and non-linearity) as the active approach but without the use of an amplifier.

\begin{equation}
\tilde{A} = \begin{bmatrix}
       \mu[1,1] \prod_{i=1}^{n} k[1,i] & \mu[1,2] \prod_{i=2}^{n} k[1,i]  & \ldots & \mu[1,n]k[1,n]         \\[0.3em]
      \mu[2,1] \prod_{i=1}^{n} k[2,i] & \mu[2,2] \prod_{i=2}^{n} k[2,i]  & \ldots & \mu[2,n]k[2,n]         \\[0.3em]
       \vdots          & \vdots & \vdots & \vdots &   \\[0.3em]
 \mu[m,1] \prod_{i=1}^{n} k[m,i] & \mu[m,2] \prod_{i=2}^{n} k[m,i]  & \ldots& \mu[m,n]k[m,n]         \\[0.3em]
     \end{bmatrix}
     \label{equationmatrix}
\end{equation}

Performing this operation on $m$ rows in the matrix $A$ results in a non-ideal matrix operation $y=\tilde{A}x$, where $\tilde{A}$ is the actual matrix in Equation \ref{equationmatrix} due to gain and droop errors. Note that since we can embed the analog non-idealities into the matrix $\tilde{A}$, the matrix operation $\tilde{A}x$ is yet another matrix operation and therefore still linear with $x$. There are a couple of approaches to ensure that the system performance is robust to this linear error. First, we ensure that $\tilde{A} \approx A$ at the $3$b level. In our $3$b MAC, $C_2$ is roughly $39\times$ larger than the maximum $C_1[i]$. Second, we show that matrix factorization \cite{lee2015factorization} allows the ability to correct for the matrix error $\tilde{A}-A$. To summarize, these non-idealities contribute errors in the matrix $A$ for both active and passive approaches but do not present non-linear effects for the passive approach. And for low-resolution multiplication (e.g. 3-bits), the errors $\tilde{A}-A$ can be either corrected for if necessary or simply ignored as long as $C_2 \gg C_1$.  Finally, the passive approach is not only more area and energy-efficient than active but also has bandwidth advantages as well. While the speed of the active approach is limited by both the settling performance of the amplifier in feedback and the RC-settling of the switches, the speed of the passive approach is set only by the RC-settling of the switches.

\subsection{Circuit Implementation}

For the reasons mentioned in Section \ref{activeVpassive}, we implement our MAC core in the SCMM using the passive apporach. The full SCMM implementation using the MAC is shown in Figure 3 where $C_{\text{s}}$ is $C_1$ and $C_{\text{ASAR}}$ is $C_2$ from Section \ref{activeVpassive}. 
We apply the SCMM for applications where the input to the MAC is both analog (application 1) and all digital (application 2). Unlike application 2, in application 1 we evaluate the MAC and SAR for use in an environment where the input is inherently analog, where the 6b input DAC is used as an on-chip test source. The MAC uses 300aF fringe capacitors for $C_{\text{s}}$. The SAR ADC uses top-plate sampling with its 6b cap-DAC $C_{\text{ASAR}}$ and digitizes the bit codes asynchronously to efficiently fit all bit decisions in the narrow time window. The 6b input DAC which generates $V_{\text{in}}[i]$ is a 5b cap-DAC with a sign-bit and is shown in Figure 3(b). The digital control signals are shown in Figure 3(c). These signals, which are preloaded from local memory, are synchronized with the rising edge of global overlapping clocks $\phi_1$ and $\phi_2$. 

After digitization, the charge on $C_\text{ASAR}$ is dumped and reset, and the operation restarts to process the next input vector. One complete inner product involves 64 cycles of sequential multiplies and accumulates using the input DAC and MAC and 1 phase of digitization by the SAR ADC.  Unlike \cite{zhang201518}, the accumulate operation is also performed in the analog charge domain, which fundamentally reduces the number of A/D conversions and rate by $64\times$ for every set of 64 multiply-and-accumulates. Figure 4 shows the chip boundary during testing and the complete compute-memory engine. The host (computer) inputs digital codes for matrices $A$ and $x$ that are loaded into local memory. Once loaded, a start token triggers the read signals to fetch from local memory. 

The 6b input DAC generates a voltage $V_{\text{in}}[i]$ to be sampled during $\phi_1$, where this voltage is prepared during the $\phi_2$ phase. During the $\phi_2$ phase, the 6b memory for the input DAC prepares $V_{\text{in}}[i]$ by either
recharging the input\textquotesingle s cap-DAC to $V_{\text{DD}}$ or discharging the previous cycle\textquotesingle s charge using $\phi_{2,\text{refresh}}$ signals. This resulting sampled charge $Q_{\text{input}}[i+1]$ generates $V_{\text{in}}[i+1] = \frac{Q_{\text{input}}[i+1]}{C_{\text{DAC,tot}}+C_\text{s}[i+1]}$ where the input DAC\textquotesingle s cap-DAC is 35$\times$ larger than $C_\text{s}[i+1]$ such that $V_{\text{in}}[i+1]$ is independent of $C_\text{s}[i+1]$ at 6b. Furthermore, when two adjacent input words share common bit codes (e.g. MSB=1 for time $i$ and $i+1$), the corresponding bit capacitor\textquotesingle s remaining charge from the previous cycle is recycled for the next input.

All the operational phases and timing are displayed in Figure 5. During  $\phi_1$, the input DAC generates a voltage that is sampled by the cap-DAC of the MAC, $C_s[i]$, which now holds $V_{\text{in}}[i] C_\text{s}[i]$. During $\phi_2$, this charge is redistributed onto $C_{\text{s,tot}}$ and $C_{\text{ASAR}}$. As stated earlier in Section \ref{activeVpassive}, $C_{\text{ASAR}} \gg C_{\text{s,tot}}$, and thus most of the sampled charge $V_{\text{in}}[i] C_\text{s}[i]$ is pushed to $C_{\text{ASAR}}$. The residual charge that remains un-transferred is at most 2.7$\%$ of the total
multiplied charge from cycle to cycle and presents a signal-independent and
correctable gain error.  While this happens, the input DAC\textquotesingle s cap-DAC prepares for the next cycle. After 61 phases of $\phi_1-\phi_2$ MAC cycles, $\phi_3$ is triggered which starts the MSB conversion of the digitization process. Input charge packets during the last few cycles only affect the last LSB SAR decisions. This prestart shaves 4 cycles or 6.3$\%$ of the total timing budget with insignificant bit errors. The SAR’s asynchronous logic finishes its LSB decision near the end of the 64th cycle. The SAR loop uses a non-monotonic binary search algorithm \cite{tripathi20138}. After the SAR algorithm, the $6$b output value is sent off-chip to be read by the host computer, and the $C_{\text{ASAR}}$ is reset to start the next inner product.

The memory is designed with decoupled write and bitline-less read because the memory read is more active than memory write. For application 1, the weights do not change at all per filter operation across the image. The idea that the weights are static is exploited in many other machine learning systems \cite{jeon201523mw} where the weights are pretrained offchip. The bitline-less, local read design is also motivated by fast read times,
obviating the need for bitline-precharge and sense-amps. For application 2, the  number of overwrite events is less than that of the read by up to 76$\%$ in certain write transitions. The local memory preloads the data at a half-cycle before each
multiply and accumulate, allowing for sufficient setup times at 2.5GHz. The codesign judiciously constrains the total memory read energy to be well-balanced with the compute energy contribution. 

\subsection{Effect of MAC's incomplete charge transfer}\label{charge}
In Section \ref{activeVpassive}, we show that the charge accumulation for passive MAC is not perfect and results in a linear and correctable error. Recall that incomplete charge transfer transforms the originally intended matrix $A$ to a matrix $\tilde{A}$ in Equation \ref{equationmatrix}. In this section, we quantify this error for our circuit implementation and propose a correction algorithm to alleviate this error whenever necessary.

As demonstration, we multiply a matrix $A \in \R^{64\times64}$ by a vector $x$ to form $y_1$, $y_2$,$\dots$,$y_{64}$. We design $A$ such that $y_1$ is highly-correlated with $x$ while $y_i$ for $i\ne1$ are uncorrelated with $x$. In Figure 6, we illustrate the ideal and actual (passive MAC) running sums of  $y_1$, $y_2$, and $y_3$. Note that what is important is the final value $y[i=64]$. We also expect that $y_1$ should reach a high value while channels $y_i$ for $i\ne1$ should be roughly $0$ at $i=64$ as illustrated in Figure 6(a). The output $y_1$ clearly accumulates in time compared to $y_2$ and $y_3$ for both ideal and passive. For the passive approach, we see attenuation that presents significant attenuation over time but which nonetheless preserves its relative value at the end of the operation at $i=64$. This is the result of incomplete charge transfer. Figure 6(b) normalizes the running sums of the passive such that $y_1[64](\text{ideal})=y_1[64](\text{passive})$. As illustrated, passive MAC attenuates all the final values $y_1[64],y_2[64],y_3[64]$ approximately equally and therefore presents an absolute gain error and not a relative error. We can nonetheless correct for this linear error using factorization by applying another matrix $B$ to invert these errors. Formally, we solve for $B$ in 
\begin{equation}
\label{eq:factor}
\begin{array}{ll} 
\mbox{minimize} & ||A-B\tilde{A}||_F \\
\mbox{subject to} & B \in \Omega_B \\
\end{array}
\end{equation}
where $F$ is the Frobenius norm, $A$ is the intended ideal matrix, $\tilde{A}$ is the actual matrix due to incomplete charge accumulation, $B$ is the correction matrix, and $\Omega_B$ is the set of possible values that $B$ can take on. For example, if the matrix $B$ were performed in fixed-point, $\Omega_B$ would contain a finite, discrete set of values allowed by fixed-point multiplication. Furthermore, cascades of matrix operations are naturally present in many applications that include AD-MM and neural networks. By applying $B\in\R^{8\times64}$ found using the algorithm in Equation \ref{eq:factor}, we obtain the corrected matrix multiplies $y_1,y_2,y_3$ as shown in Figure 6(c) and the error between ideal and corrected in (d). It is important to note that the performance of factorization is a strong function of the aspect ratio of $B \in \R^{m\times n}$.  Figure 7 illustrates the matrix reconstruction error $||A-B\tilde{A}||_F$ for various $m$ keeping $n$ constant. 

We also illustrate the effect of incomplete charge transfer from the perspective of the matrix itself. Figure 8(a) illustrates the ideal matrix $A$, and Figure 8(b,c) illustrate the uncorrected matrix $\tilde{A}$ and calibrated matrix $B\tilde{A}$. For this simulation, we accentuate the amount of attenuation for qualitative effect by setting $C_{\text{ASAR}}$ to be only $10\times$ larger than the maximum $C_{\text{s}}$ instead of $39\times$ in our actual implementation. 

\subsection{Noise Analysis of MAC}
The MAC system in Figure 3 is designed such that its output noise is well within the 6b SNR target of the ADC. In spite of the $300$aF unit capacitors, the thermal noise contributions from the MAC, the input DAC's cap-DAC, and asynchronous SAR (comparator) are designed to be well below the least significant bit (LSB) (7mV) of the ADC's output. 
 
Since the MAC's total capacitance is $35\times$ smaller than that of the input DAC, we detail only the $kTC$ noise of this block due to its much greater contribution of thermal noise. The noise contribution from the MAC's cap-DAC is as follows. Let the noise voltage (RMS) on $C_\text{ASAR}$ be $\sigma_{C_\text{ASAR}}[i]$ at time $i=1,\dots,64$. During the first cycle, $\phi_1$ is turned on and off. This generates a sampled noise charge $kTC_\text{s}[i]$ on $C_\text{s}[i]$. When $\phi_2$ is turned on, this charge is redistributed onto $C_\text{s,tot.}$ and $C_\text{ASAR}$. The noise voltage across $C_\text{ASAR}$ is then $\frac{kTC_\text{s}[i]}{(C_\text{s,tot.}+C_\text{ASAR})^2}$. When $\phi_2$ is subsequently turned off, this generates another sampled noise charge on $C_\text{ASAR}$ of   $kT(\frac{C_\text{s,tot.}C_\text{ASAR}}{C_\text{s,tot.}+C_\text{ASAR}})$. The total noise voltage on $C_\text{ASAR}$ after a complete $\phi_1$-$\phi_2$ cycle is $ \sigma_{C_\text{ASAR}}[i=1]  = \sqrt{\frac{kT|C_\text{s}[i]|}{(C_\text{s,tot.}+C_\text{ASAR})^2} + \frac{kT(\frac{C_\text{s,tot}C_\text{ASAR}}{C_\text{s,tot}+C_\text{ASAR}})}{C_\text{ASAR}^2}} $.

After $i$  $\phi_1$-$\phi_2$ cycles, the noise voltage on $C_\text{ASAR}$ becomes
\begin{eqnarray}
\label{eq:wandx}
\sigma_{C_\text{ASAR}}[i] & = \sqrt{\left( \frac{kT|C_\text{s}[i]|}{(C_\text{s,tot.}+C_\text{ASAR})^2} + \frac{kT\left(\frac{C_\text{s,tot}C_\text{ASAR}}{C_\text{s,tot}+C_\text{ASAR}}\right)}{C_\text{ASAR}^2} \right) \left(\sum_{j=0}^{i}\left( \frac{C_{\text{ASAR}}}{C_{\text{ASAR}}+C_{\text{s,tot.}}}\right)^{2j}\right)} 
\end{eqnarray}
 
 Due to the time-varying nature of $C_{\text{s}[i]}$, it is rather difficult to gain any intuition. We can however conclude that the minimum noise occurs when  $C_{\text{s}[i]}=0$ and the highest occurs when  $C_{\text{s}[i]}=C_{\text{s,tot.}}$. At the highest noise power, we simplify this expression as

\begin{eqnarray}
\label{eq:brown}
\sigma_{C_\text{ASAR}}[i] = \sqrt{\frac{kT}{C_\text{ASAR}}\frac{C_\text{s,tot}(C_\text{s,tot}+2C_\text{ASAR})}{(C_\text{s,tot}+C_\text{ASAR})^2} \sum_{j=0}^{i}\left( \frac{C_\text{ASAR}}{C_\text{s,tot}+C_\text{ASAR}}\right)^{2j}}
\end{eqnarray}

Interestingly, for many MAC cycles $i\to\infty$, the noise power reaches equilibrium (or steady-state) and simply becomes $\lim_{i\to\infty} \sigma[i]  = \sqrt{ \frac{kT}{C_{\text{ASAR}}}}$. However, for our implementation the MAC cycles stop at time $i=64$; therefore, the noise due to the MAC is slightly lower than $\sqrt{ \frac{kT}{C_{\text{ASAR}}}}$, the steady-state value. 

To validate this, we perform Monte-Carlo transient noise simulations in 40nm for the MAC circuit during the $64$ $\phi_1-\phi_2$ multiply and accumulate phases and where we set $C_{\text{s}}[i] = C_{\text{s,tot.}}$. Figure 9 plots $150$ Monte-Carlo transients for the differential output voltage on $C_{\text{ASAR}}$ and the predicted $3\sigma$ line from Equation \ref{eq:brown}. The simulated variance grows with the number of MAC cycles and shows close agreement to analytical predictions. We also highlight that the actual thermal noise of the MAC is less than $\frac{kT}{C_{\text{ASAR}}}$ during the MAC's operating window.

\section{Chip Measurements and Applications}
The SCMM was fabricated in 40nm CMOS, and its die photo is shown in Figure 10. The performance of matrix multiplication $y=Ax$ is measured, $A\in\R^{8\times64}$ contains orthonormal row vectors where each row of $A$, denoted by $a^T_j$, is orthogonal to all $a^T_k$ where $j \ne k$, and $x=a^T_l$ for $l \in \{1,\dots,8\}$. Many applications arise where the matrix operation is an orthonormal transformation. These include Discrete Fourier and Cosine Transformations (DFT, DCT) and matched filtering. Here we apply such transformations to characterize matrix multiplication performance.

We measure $y_j=Ax^{(i)}$ where $x^{(i)} = a^T_i$ for $8$ trials $i\in \{1,\dots,8\}$ and for $8$ output channels $j\in \{1,\dots,8\}$ using a clock rate of 1GHz. Note that the $x^{(i)}$ is a vector, which is different than the scalar element $x[i]$, which denotes the value at time $i$. Ideal multiplication result is shown in Figure 11(a), and the measured output is shown in Figure 11(d) as raw digital codes and normalized in Figure 11(b). Note that for clarity, we have omitted the quantization operation that occurs when sending in the values for $A$ and $x$ to memory. Figure 11(a) indicates that under ideal matrix operation $Ax$, we expect to see $y_j = 1$ for $j=i$ and $y_j=0$ for $j\ne i$. However, due to quantization of $A$ and $x$ as well as non-idealities in the circuit operation, we observe Figure 11(b), which clearly shows non-zero values of $y_j$ for $j\ne i$. The mean square error (MSE) of the ideal to the measured is $0.00723$. Using matrix factorization as described in Section \ref{charge}, we obtain Figure 11(c), which attenuates the cross-terms $y_j$ for $j\ne i$ and equalizes the diagonal terms $y_j$ for $j = i$. 

We measure the system noise performance using a matched filtering setup at 1GHz. Here, we perform a single inner product $y=\sum_{i=1}^{64}a[i]x[i]$ where $x=a+n$, $a$ is a chirp signal, and $n$ is independent and identically distributed additive white gaussian noise source. We vary the input SNR by sweeping the variance of $n$ and plot the output response $y$ in Figure 12(a) and perform $25$ independent trials. The output voltage is derived from using the output codes and the LSB size (7mV) of the ADC outputs. We plot the mean and variance of the output for varying levels of input SNR. Increasing input SNRs decreases the noise contribution from the input source (variance) until around $5$ dB input SNR when the output variance drops to the noise floor set by the LSB size. The system offset in Figure 12(b) is similarly obtained by measuring the mean of the inner products of random vectors, which ideally converges to $0$. The offset of the entire system due to leakage,
comparator, and capacitor mismatch contributions is within half-LSB of the ADC output at
speeds $>0.5$GHz.

In our first application, we test our MAC implementation for use in a neural network image classifier. Image classifiers with convolutional neural networks use a cascade of linear matrix multiplies followed by non-linearity functions. After training, the first layer of a network contains Gabor-like features that are used to detect edges. Figure 13 illustrates the ideal activations in (c) when 1 Gabor-like filter shown in (b) is scanned over the red channel of the colored input image in (a). This scanning process is performed by multiplying the input image with the filter for varying horizontal and vertical offsets. We obtain results from our chip (d) and simulated digital (e,f). The simulated digital (e) uses fixed-point arithmetic with 6b input, 3b weights, and 6b output notated as (6b/3b/6b), but with infinite accumulator resolution while (f) uses 6b/3b/6b with 6b accumulator. 

We extend this primitive to reduce the A/D footprint when the first neural network layer operates on inherently analog inputs. For systems that process inherently analog inputs, multiplying a full-rank matrix $A$ of size $m \times n$ in the analog domain reduces the A/D rate by $n$ and compresses the total number of A/D conversions by factor $n/m$ when $n>m$ where analog matrix multiplication is made practical using factorization. We extend this idea by applying this compression technique to a classifier in a neural network. This front-end layer both compresses and classifies analog image data in an end-to-end pretrained neural network as  illustrated in Figure 14. To reduce the dimensionality, the matrix-vector operations are processed at 1GHz in non-overlapping regions of the image with a stride equal to the filter width and height of size 8 and output dimension smaller
than its input dimension. Three filters are applied per color channel, and the
resulting activations of size $4\times4\times9$ are digitized and pipelined to the remaining
two layers in the digital domain. The dimensionality of the output of this layer is $4^2 \times 9$ while that of the input is $32^2\times3$; this layer therefore compresses the data by a factor of $21.3\times$ and furthermore decreases the number of A/D operations by $21.3\times$ and digitization rate by $64\times$.  

This front-end layer and digital layers are cotrained
together in floating-point using real images from the Cifar-10 database
\cite{krizhevsky2009learning}. The performance summary is shown in Table 1. The reduced-precision digital layer using digital fixed-point yields 86$\%$ top-3 accuracy while the proposed analog layer yields 85$\%$ top-3 accuracy. The digital fixed-point layer is simulated in Matlab and uses $6$ bits for input, $4$ bits for weights, and $6$ bits for outputs with full-precision accumulation and no overflow. The energy per operation of the proposed analog layer is $11\times$ less than that of digital, which is simulated in $40$nm. The synthesized digital MAC uses $6$b input, $4$b weights, and $6$b outputs with a $6$b accumulator. Note that to prevent overflow, digital MACs are designed to have an output bitwidth that is at least the sum of the bitwidths of the input and weights. More commonly, the output bitwidth is set to be  $\log_2(n)$ larger than the sum of the input bitwidths, where $n$ is the number of MAC products. However, for our energy comparisons, we are conservative in our digital energy estimation, and as such we set the digital output bitwidth to be $6$b instead of $6+4+\log_2(64)=15$ bits. Finally, the proposed analog layer lowers the number of A/D conversions by $21\times$ as compared to conventional digital approach. The compute to memory read energy ratio is 1.18:1.


 For our second application, we demonstrate analog co-processing and acceleration for computation that is traditionally performed in the digital domain.  This application computes gradients in Stochastic Gradient Descent, one of the most widely used optimization algorithms in high performance computing. The goal of the
optimization algorithm is to find $\theta$, the vector of unknown parameters that finds
the global optimal solution by minimizing a user-defined objective function $J(\theta)$, which
is non-convex in general. Here, the SCMM is used to maximize  test accuracy for image classification by minimizing $J(\theta)$ over a training set by iteratively updating $\theta^{(i+1)} := \theta^{(i)} + \alpha \nabla J(\theta^{(i)})$, where $i$ represents the iteration count, $ \nabla J(\theta^{(i)})$ the gradient evaluated at $\theta^{(i)}$ from a sampled batch, and $\alpha$ the learning rate. Evaluating the gradients $\nabla J(\theta^{(i)})$ is an expensive operation that usually consists of a large matrix multiply for many classifiers including neural networks. Here we perform gradient descent on an image classification task and offload the gradient computation $\nabla J(\theta^{(i)}) = A^{(i)}[x_1^{(i)}, \dots, x_m^{(i)}]$ with the SCMM chip, where $m$ is the batch size, and $x_1^{(i)}$ corresponds to the data vector from one image sample. The SCMM takes as inputs a digital matrix $A$ and a digital vector $x$, and performs the matrix
operation $Ax$ in the analog charge-domain. The 6b output
is the gradient update at each iteration. The 100 gradient updates are performed
on a sample classification problem with a learning rate $\alpha=10^{-6}$. The accuracy of our chip is compared with various simulated digital computations in Figure 15.  The measured solution to the optimization problem is shown in Figure 16, which shows close alignment with simulated digital. 
Analog acceleration using the SCMM at 2.5GHz performs slightly worse than digital
double-precision 64b and is equivalent to simulated digital fixed-point at an
estimated $6\times$ lower energy and the compute to memory read energy ratio is 1.05:1.
 
Table 2 summarizes the performance of the analog charge-domain MAC for two applications as compared with a recent work of embedding multiplication in a SAR ADC \cite{zhang201518}. The efficiencies are computed based on measured power and speed. The measured efficiency including compute, memory, self-test logic, and clock is 8.7TOPS/W at 1GHz for application 1, and 7.7TOPS/W at 2.5GHz for application 2. As compared to  \cite{zhang201518} (130nm), our work (in 40nm) is $2$ orders of magnitude more energy-efficient, where our energy also includes the DAC, clocking, and memories. However, it is worth mentioning that \cite{zhang201518} embeds digital barrel-shifting with analog fixed-point multiplication, which allows higher multiplication resolution but which makes analog-domain accumulation difficult to realize.
  
\section{Acknowledgements} \label{acknowledgements}

The authors thank D. Miyashita, B. Murmann, M. Udell, D.
Bankman, C. Young, K. Zheng, the
TSMC University Shuttle Program, and NSFGRFP
and TI Stanford Graduate Fellowship.

\section{Conclusion} \label{conclusion}
This work presents the SCMM, which uses switches,  $300$aF unit capacitors and local memories. We present general results for multiplication and characterization of noise and offset. We also analyze MAC imperfections such as thermal noise and incomplete charge accumulation. We show that this error is linear with the input, and is correctable using matrix factorization. Finally, we demonstrate the SCMM on two applications with high efficiency at above-GHz MAC rates.

\newpage

\section{List of Figures and Tables} \label{list_of_figures}
\begin{enumerate}
\item Figure 1. Analog matrix multiplication for various signal processing applications. 
\item Figure 2. (a) Active analog MAC and (b) passive analog MAC implementations.
\item Figure 3. (a) SCMM implementation, (b) SCMM’s 6b input DAC, and (c) digital controls.
\item Figure 4. (a) SCMM chip boundary when under test and (b) compute-memory architecture.
\item Figure 5. The operational modes and timing diagram.
\item Figure 6. The ideal and simulated analog MAC transient outputs for a matrix operation $Ax$.
\item Figure 7. Matrix factorization reconstruction error (normalized MSE) as a function of $m$.
\item Figure 8. (a) A programmed 64x64 matrix $A$, (b) the matrix that is a result of incomplete accumulation $\tilde{A}$, and (c) the corrected matrix using matrix factorization.
\item Figure 9. Transient noise simulation of the noise voltage on $C_{\text{ASAR}}$.
\item Figure 10. Photograph of the chip in 40nm CMOS.
\item Figure 11. (a) Ideal matrix-vector product output, (b) measured output, (c) corrected output, (d) measured digital code. 
\item Figure 12. Measured matched filter response for varying input SNRs and the measured offset.
\item Figure 13. Measured versus simulated performance of a filter (b) on an image (a).
\item Figure 14. Designed compression and classifying layer for a small convolutional neural network and measured confidence levels for sampled input images on the CIFAR10 database.
\item Table 1. Performance of the compression layer versus conventional.
\item Figure 15.  The classification accuracy as a function of the number of gradient steps in the stochastic gradient descent procedure.  The NMSE of analog averaged over 100 steps is 0.006.
\item Figure 16. Final optimization solution after 100 gradient steps.
\item Table 2. Performance summary.
\end{enumerate}

\newpage 

\begin{figure}[]
  \centering
    \includegraphics[width=0.8\textwidth]{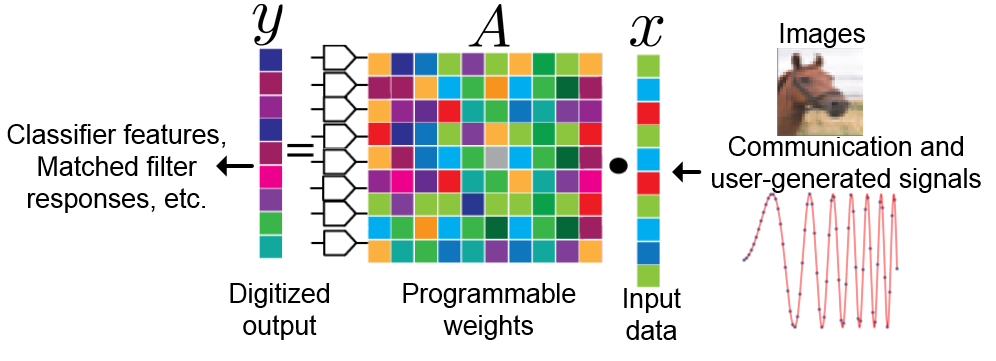}
\caption{Analog matrix multiplication for various signal processing applications. }
\end{figure}

\begin{figure}[]
  \centering
    \includegraphics[width=0.8\textwidth]{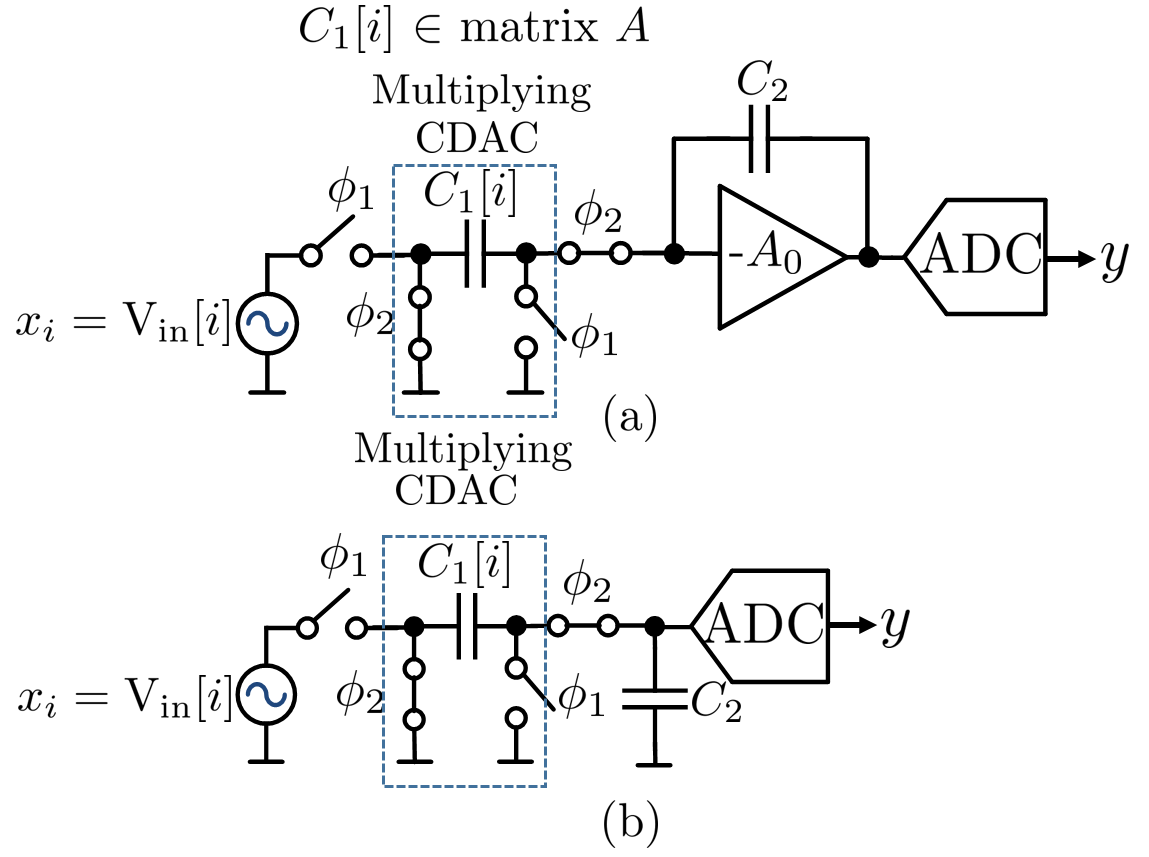}
  \caption{(a) Active analog MAC and (b) passive analog MAC implementations.}
\end{figure}

\begin{figure}[t]
  \centering
    \includegraphics[width=0.8\textwidth]{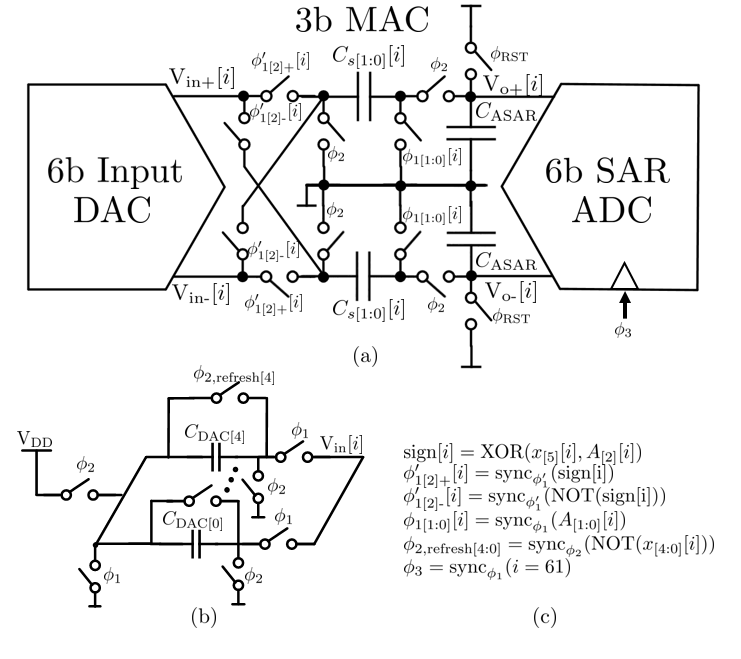}
  \caption{(a) SCMM implementation, (b) SCMM’s 6b input DAC, and (c) digital controls. }
\end{figure}

\begin{figure}[]
  \centering
    \includegraphics[width=0.8\textwidth]{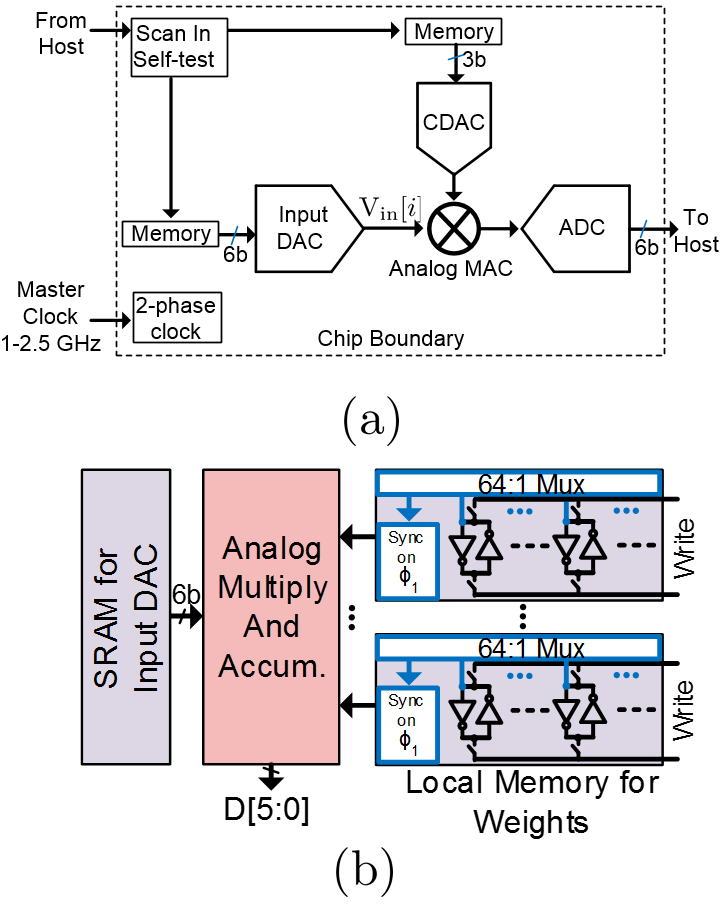}
  \caption{ (a) SCMM chip boundary when under test and (b) compute-memory architecture.}
\end{figure}

\begin{figure}[]
  \centering
    \includegraphics[width=0.8\textwidth]{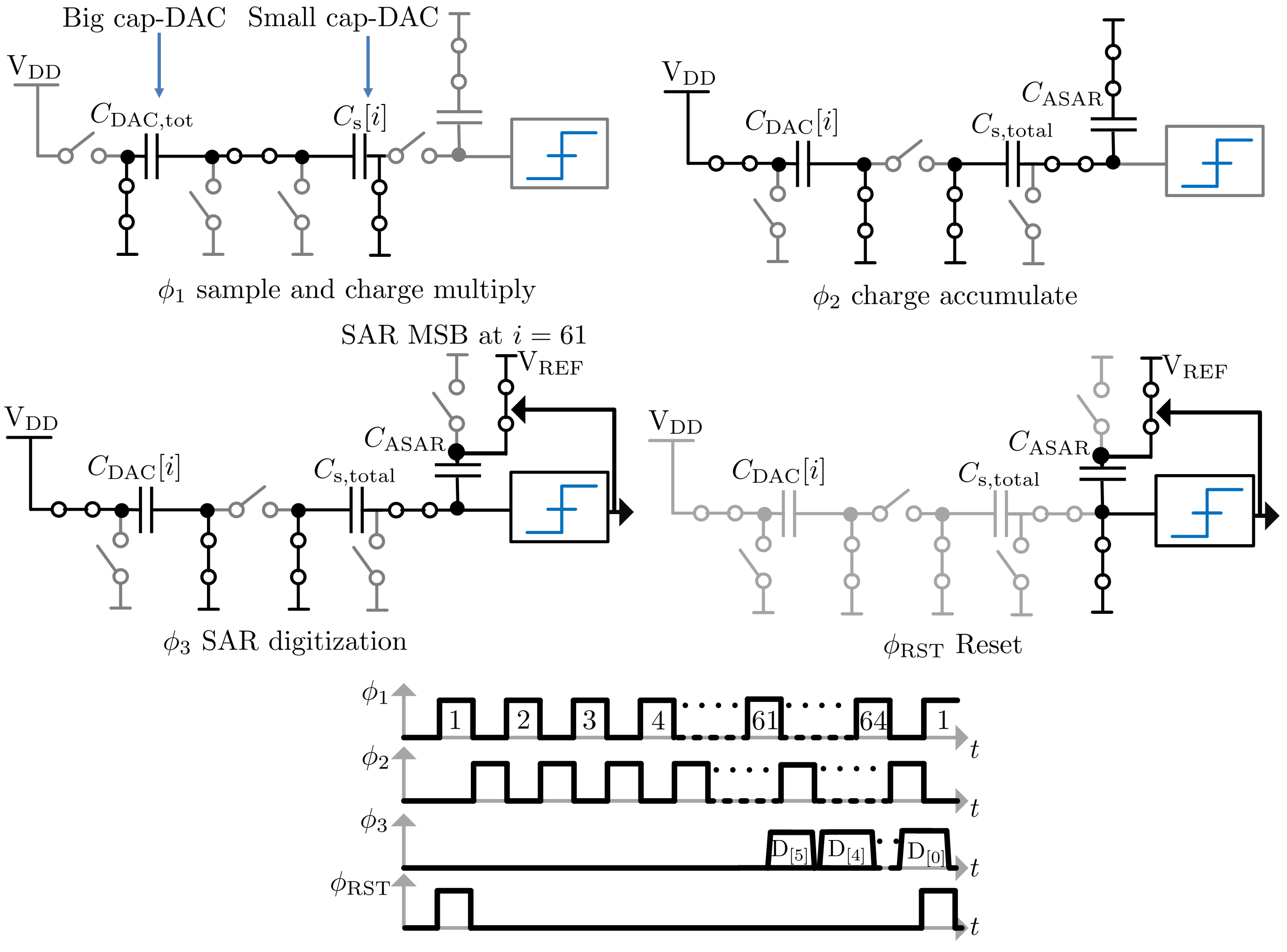}
  \caption{The operational modes and timing diagram.}
\end{figure}

\begin{figure}[]
  \centering
    \includegraphics[width=0.8\textwidth]{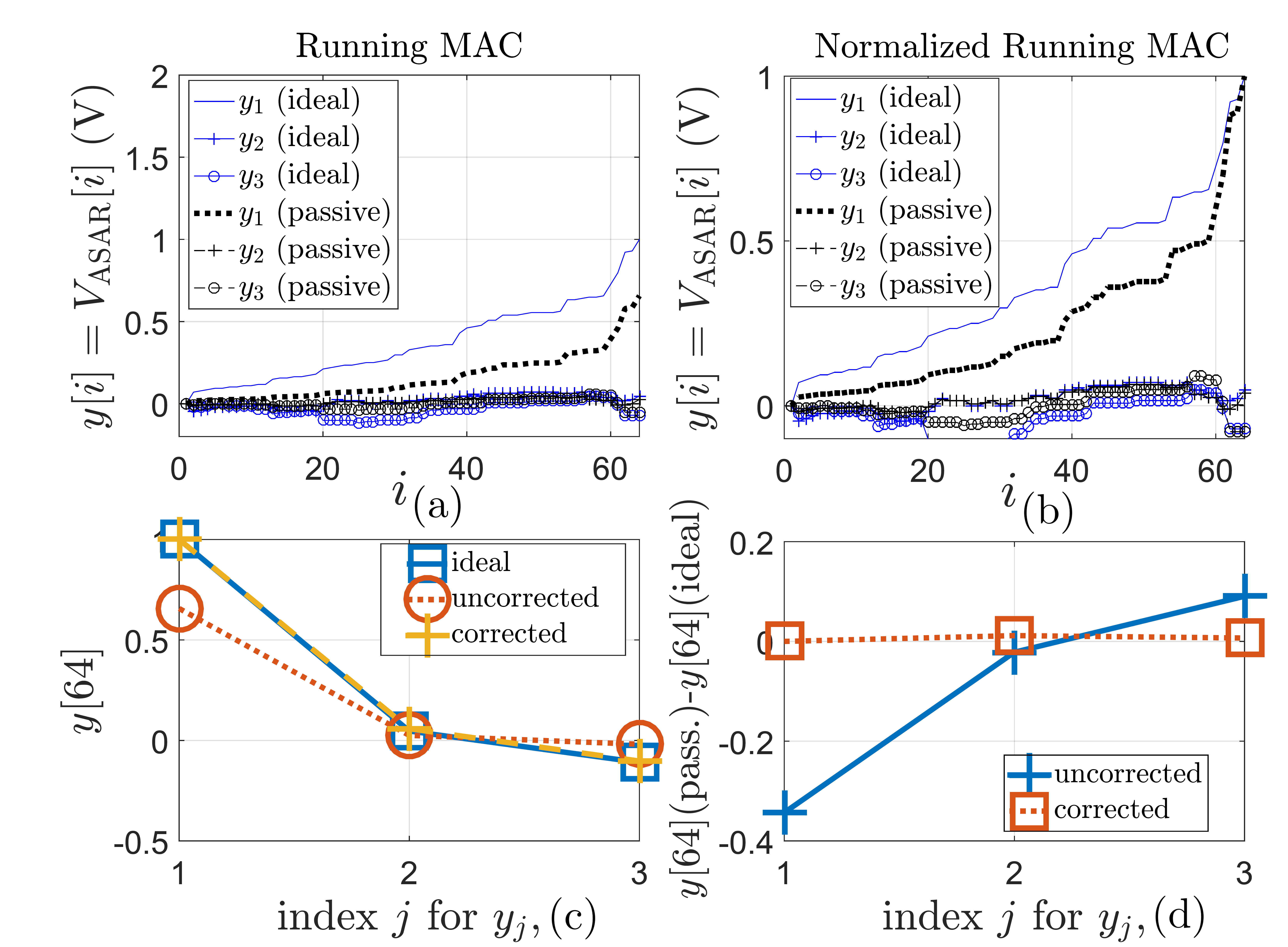}
  \caption{The ideal and simulated analog MAC transient outputs for a matrix operation $Ax$.}
\end{figure}

\begin{figure}[]
  \centering
    \includegraphics[width=0.5\textwidth]{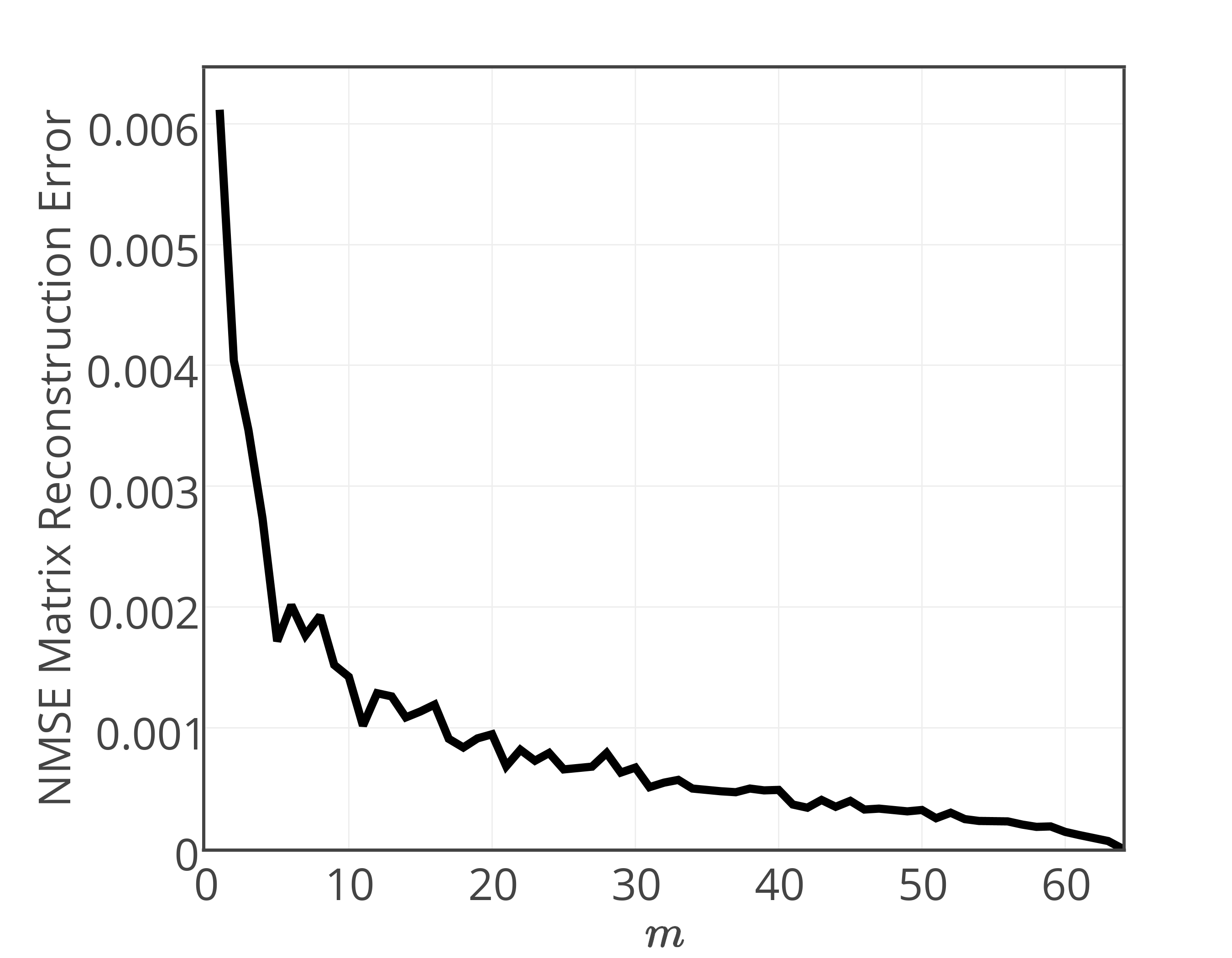}
  \caption{Matrix factorization reconstruction error (normalized MSE) as a function of $m$.}
\end{figure}

\begin{figure}[]
  \centering
    \includegraphics[width=0.8\textwidth]{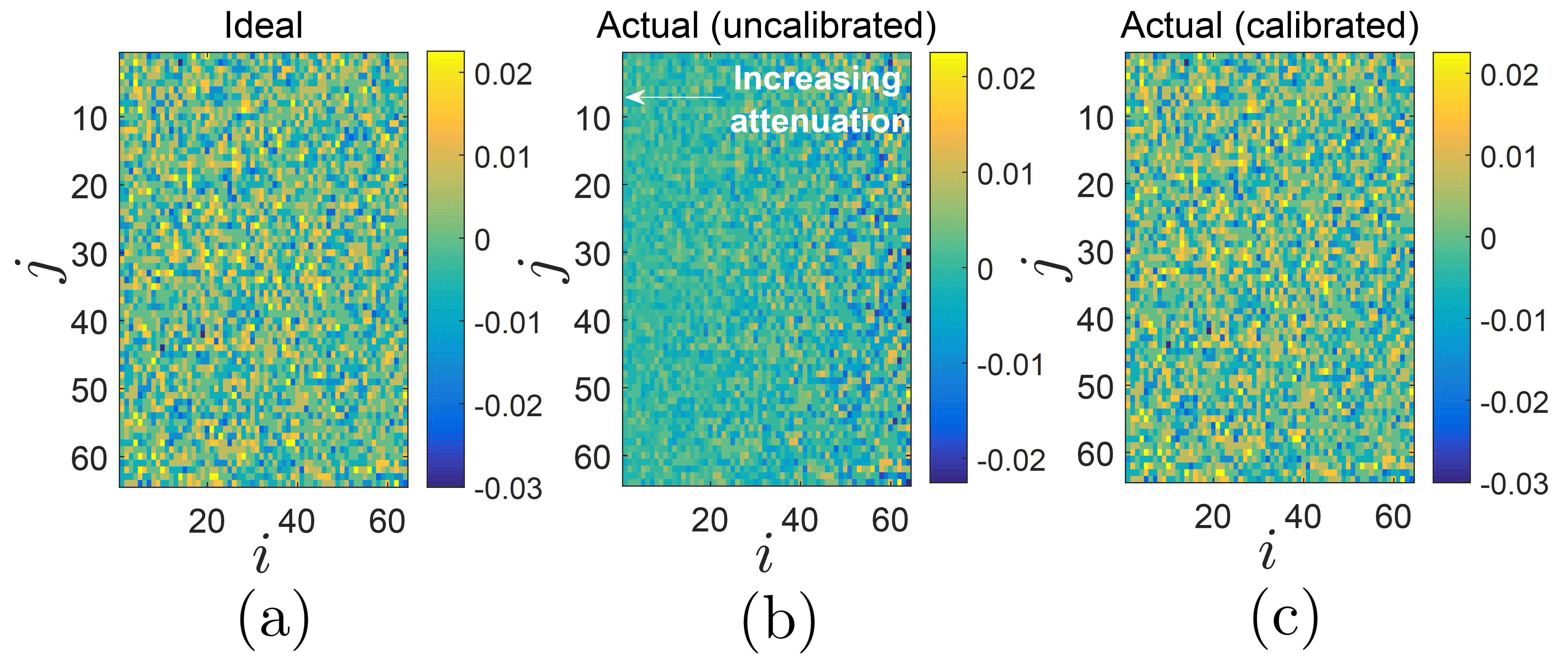}
  \caption{(a) A programmed 64x64 matrix $A$, (b) the matrix that is a result of incomplete accumulation $\tilde{A}$, and (c) the corrected matrix using matrix factorization.}
\end{figure}

\begin{figure}[]
  \centering
    \includegraphics[width=0.5\textwidth]{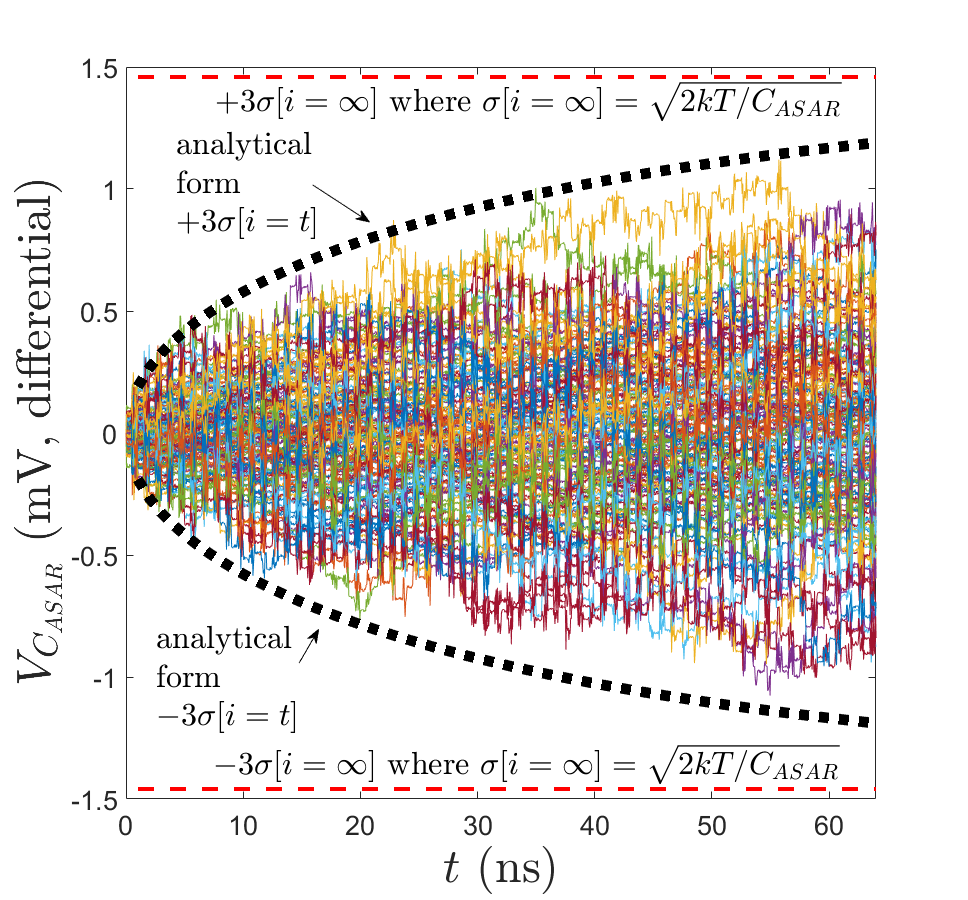}
  \caption{Transient noise simulation of the noise voltage on $C_{\text{ASAR}}$.}
\end{figure}

\begin{figure}[]
  \centering
    \includegraphics[width=0.5\textwidth]{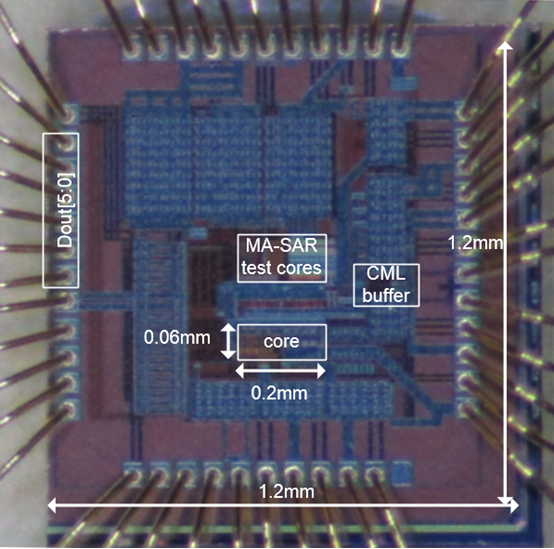}
  \caption{Photograph of the chip in 40nm CMOS.}
\end{figure}

\begin{figure}[]
  \centering
    \includegraphics[width=0.7\textwidth]{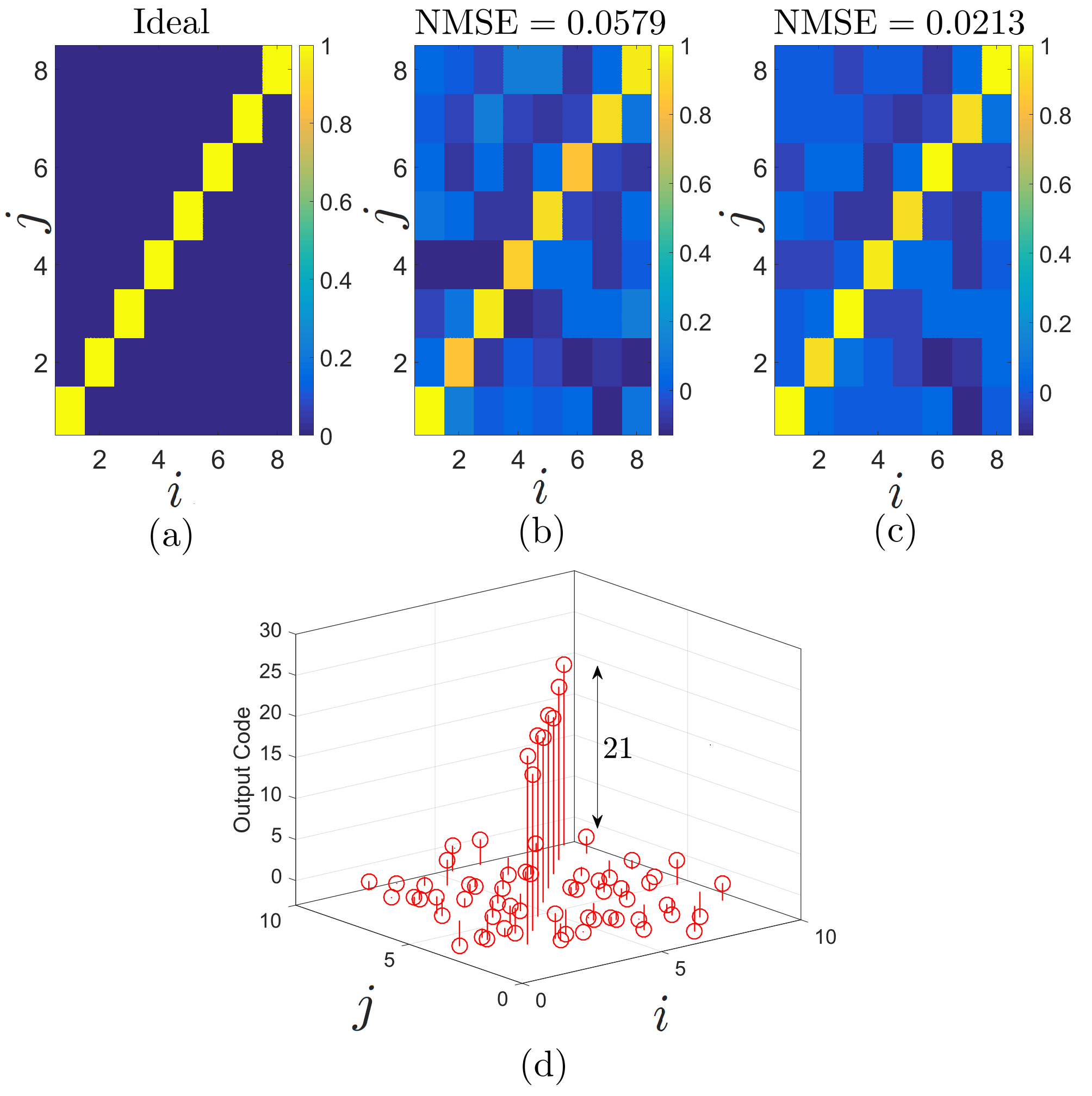}
  \caption{(a) Ideal matrix-vector product output, (b) measured output, (c) corrected output, (d) measured digital code. }
\end{figure}

\begin{figure}[]
  \centering
    \includegraphics[width=\textwidth]{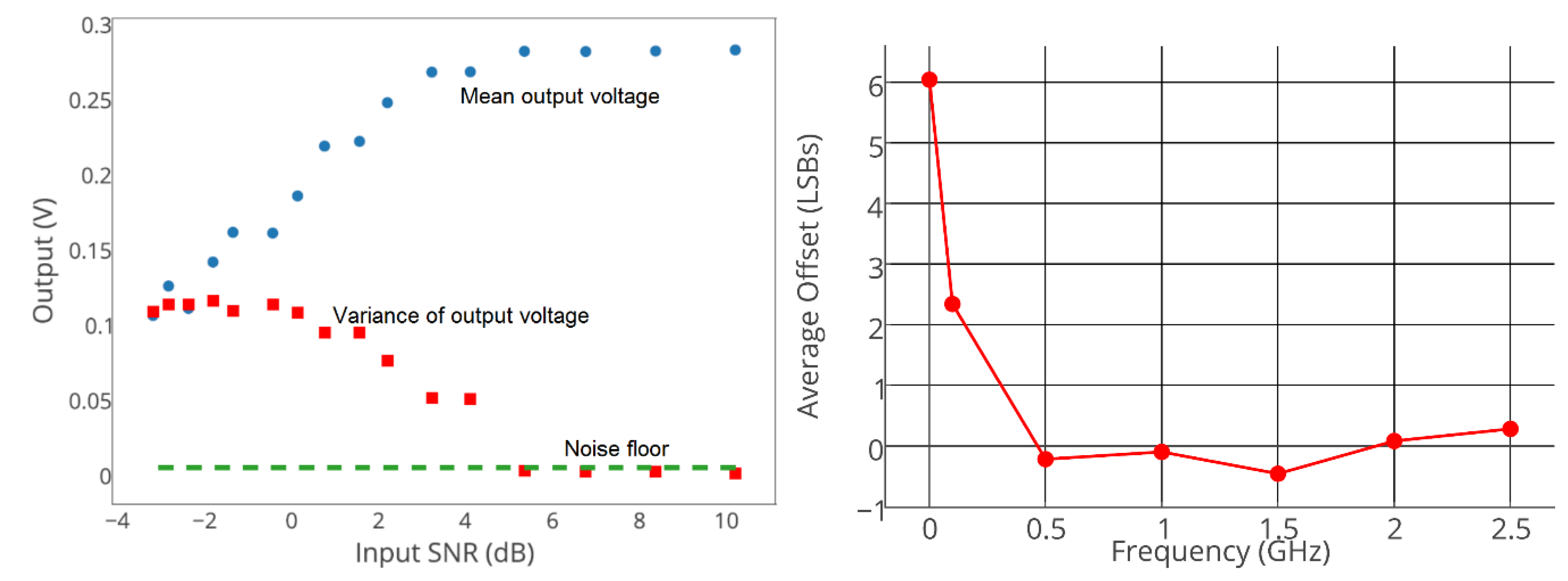}
  \caption{Measured matched filter response for varying input SNRs and the measured offset.}
\end{figure}

\begin{figure}[]
  \centering
    \includegraphics[width=0.5\textwidth]{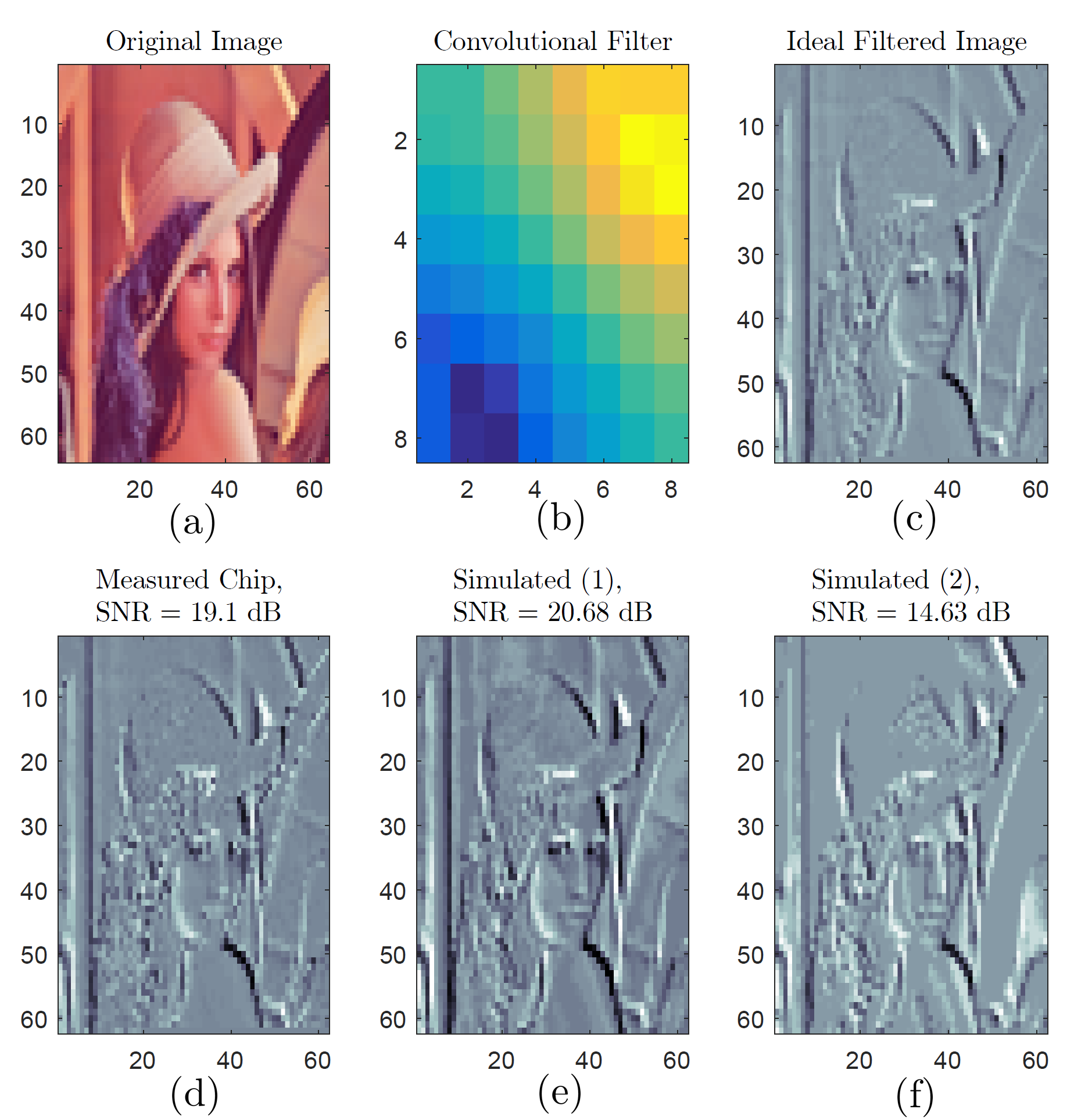}
  \caption{Measured versus simulated performance of a filter (b) on an image (a).}
\end{figure}

\begin{figure}[]
  \centering
    \includegraphics[width=0.5\textwidth]{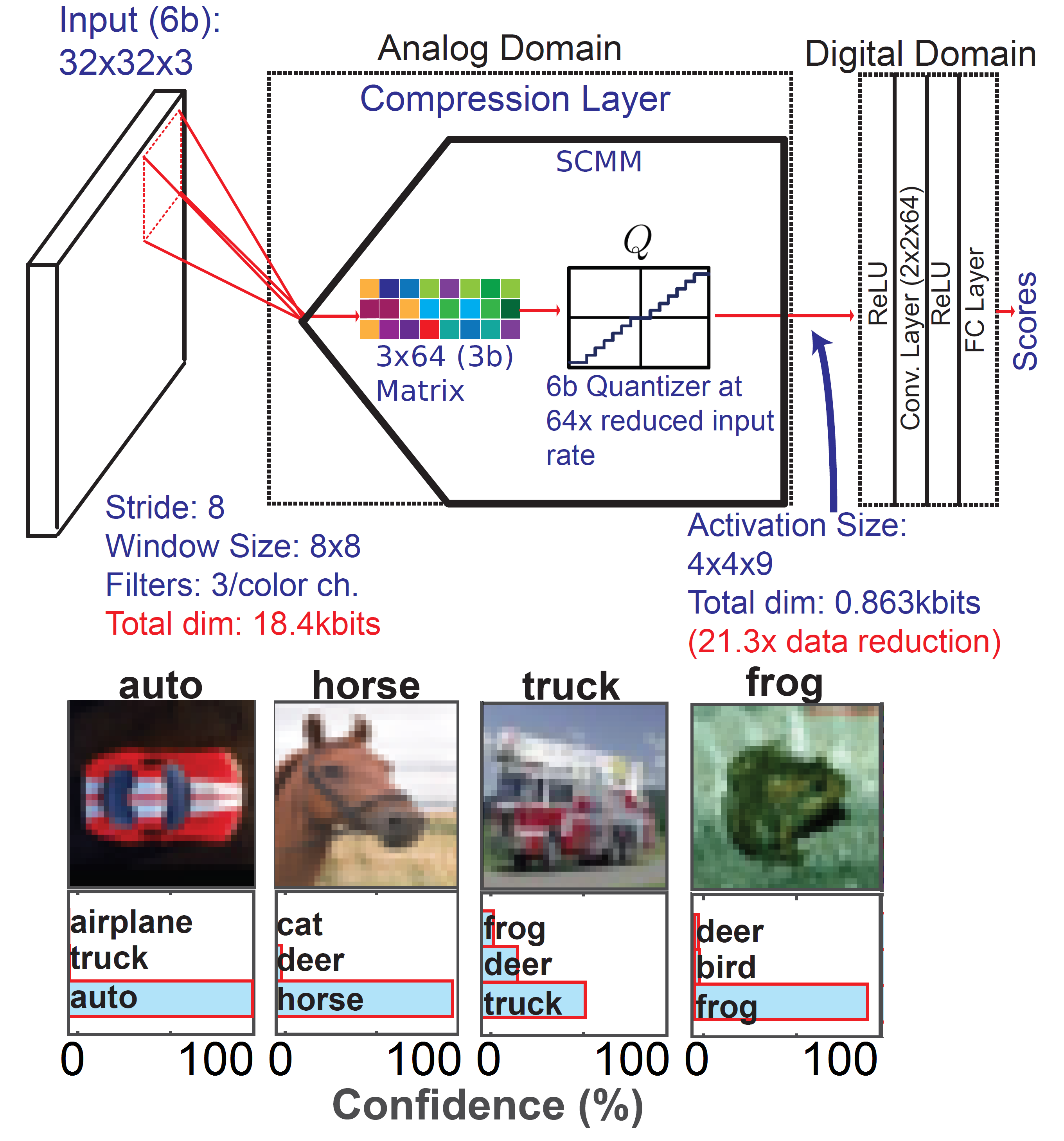}
  \caption{Designed compression and classifying layer for a small convolutional neural network and measured confidence levels for sampled input images on the CIFAR10 database.}
\end{figure}

\begin{figure}[]
  \centering
    \includegraphics[width=0.5\textwidth]{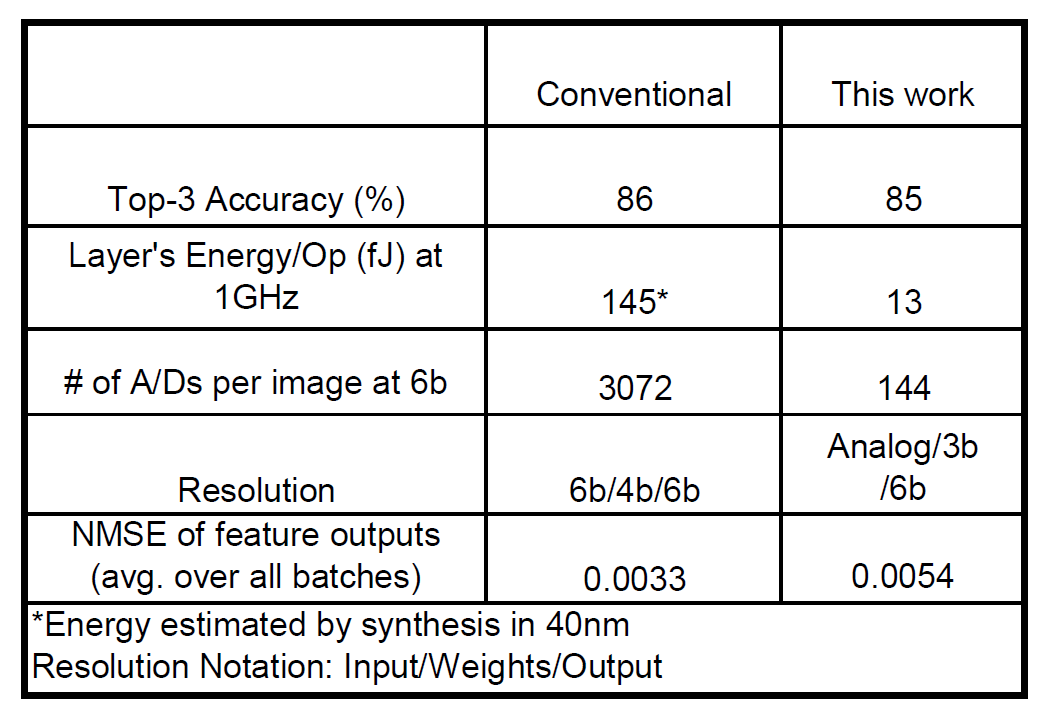}
  \caption{Table 1. Performance of the compression layer versus conventional.}
\end{figure}

\begin{figure}[]
  \centering
    \includegraphics[width=0.5\textwidth]{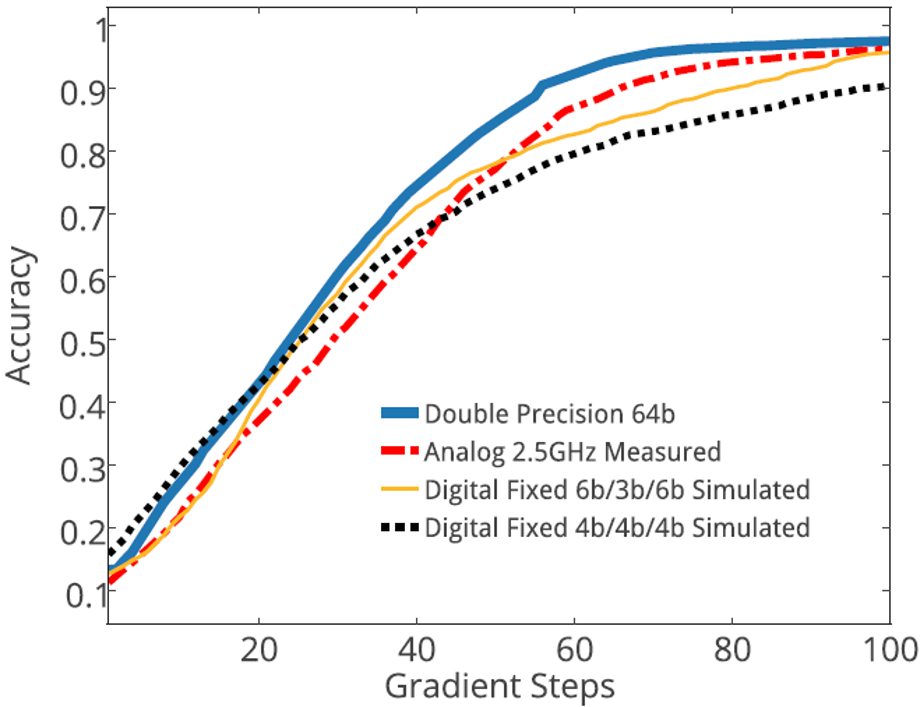}
  \caption{The classification accuracy as a function of the number of gradient steps in the stochastic gradient descent procedure.  The NMSE of analog averaged over 100 steps is 0.006.}
\end{figure}

\begin{figure}[]
  \centering
    \includegraphics[width=0.5\textwidth]{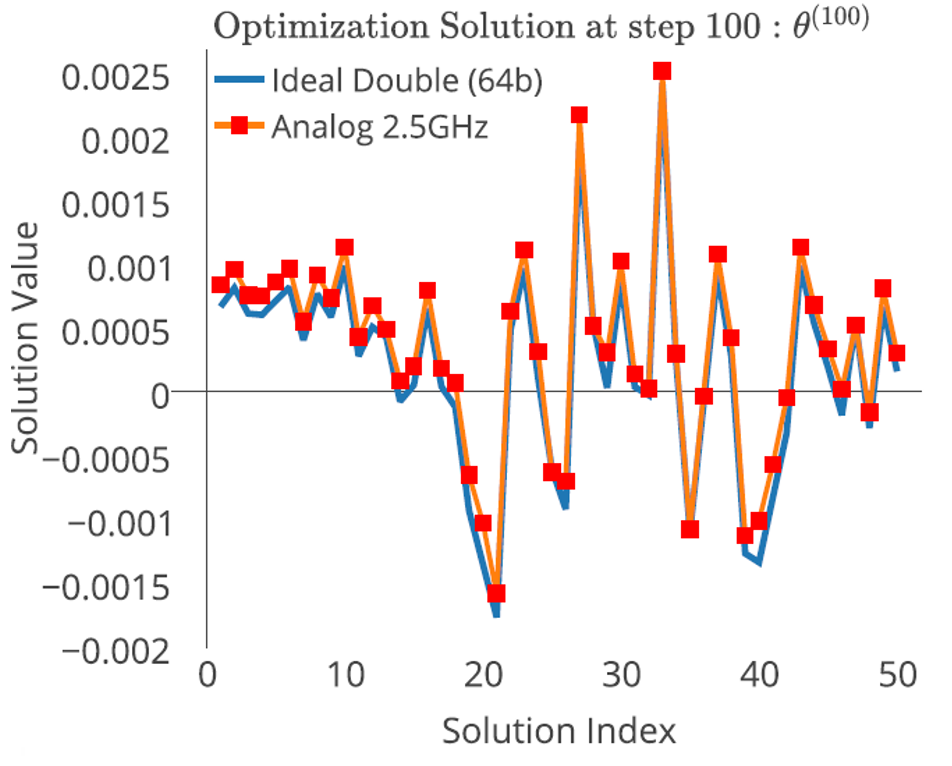}
 \caption{Final optimization solution after 100 gradient steps.}
\end{figure}

\begin{figure}[]
  \centering
    \includegraphics[width=0.8\textwidth]{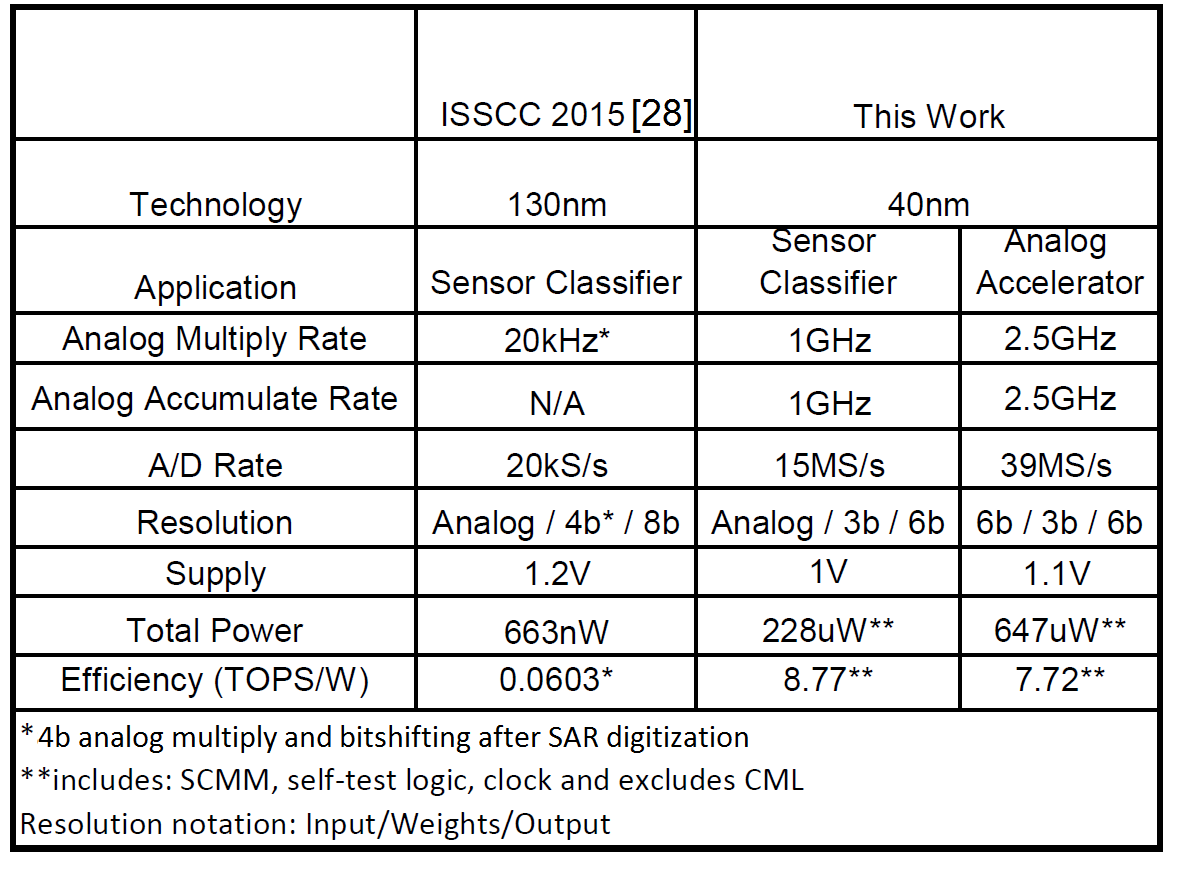}
  \caption{Table 2. Performance summary.}
\end{figure}

\clearpage

{\small
\bibliographystyle{unsrt}
 
\bibliography{egbib2}
}

\end{document}